\documentclass[12pt]{iopart}
\usepackage{graphicx}
\begin{document}

\title{Niobium-based superconducting nano-devices fabrication using all-metal suspended masks}

\author{S. Samaddar}
\address{Institut N\'eel, CNRS, Universit\'e Joseph Fourier and Grenoble INP, 25 rue des Martyrs, 38042 Grenoble, France}
\ead{sayanti.samaddar@grenoble.cnrs.fr}
\author{D. van Zanten}
\address{Institut N\'eel, CNRS, Universit\'e Joseph Fourier and Grenoble INP, 25 rue des Martyrs, 38042 Grenoble, France}
\author{A. Fay}
\address{Institut N\'eel, CNRS, Universit\'e Joseph Fourier and Grenoble INP, 25 rue des Martyrs, 38042 Grenoble, France}
\author{B. Sac\'ep\'e}
\address{Institut N\'eel, CNRS, Universit\'e Joseph Fourier and Grenoble INP, 25 rue des Martyrs, 38042 Grenoble, France}
\author{H. Courtois}
\address{Institut N\'eel, CNRS, Universit\'e Joseph Fourier and Grenoble INP, 25 rue des Martyrs, 38042 Grenoble, France}
\ead{herve.courtois@grenoble.cnrs.fr}
\author{C. B.  Winkelmann}
\address{Institut N\'eel, CNRS, Universit\'e Joseph Fourier and Grenoble INP, 25 rue des Martyrs, 38042 Grenoble, France}
\ead{clemens.winkelmann@grenoble.cnrs.fr}

\begin{abstract}
We report a novel method for the fabrication of superconducting nano-devices based on niobium. The well-known difficulties of lithographic patterning of high-quality niobium are overcome by replacing the usual organic resist mask by a metallic one. The quality of the fabrication procedure is demonstrated by the realization and characterization of long and narrow superconducting lines and niobium-gold-niobium proximity SQUIDs.
\end{abstract}
\maketitle

\section{Introduction}

Superconductivity is an essential ingredient in quantum nano-electronics. Its macroscopic phase coherence allows to produce easily large objects obeying quantum mechanics \cite{Vion2002, Wallraff04}. Aluminum is presently the emblematic material in this field. Its popularity is mainly due to the ease of its lithographic patterning and the high quality of its native oxide. The critical temperature $T_c=1.17$ K of bulk aluminum is however rather low, calling for the use of dilution refrigerators and elaborate electronic filtering as temperatures $T \ll T_c$ are usually needed. Moreover, the relevant energy scale of superconductivity, namely the superconducting gap $\Delta$, is usually proportional to $T_c$. A low critical temperature $T_c$ is therefore synonymous of a low critical current in short Josephson junctions or superconducting quantum interference devices (SQUIDs) \cite{Tinkham}.

Higher $T_c$ elemental superconducting metals such as vanadium and lead are occasionally used \cite{Spathis11}, but they suffer from rapid aging under ambient air conditions. Niobium is the highest-$T_c$ simple element (9.2 K) and is also rather immune to aging. Its main drawback is the strong sensitivity of its thin film superconducting properties to contamination. As niobium is a refractory metal, its evaporative deposition implies extremely high target temperatures. In the case of lift-off processes, this leads to significant heating and outgassing of the organic resist mask, provoking a strong reduction and scatter of the effective critical temperature $T_c$ \cite{Wei2008, Ohnishi08}. Nb nanostructures with a high $T_c$ can be fabricated within a lift-off process using a deposition set-up in which the Nb target is far away from the sample \cite{Kim01}; this method is however difficult to transpose to most setups.

Several techniques have been used to circumvent this severe limitation to niobium applications. Sputtering can provide niobium films of high quality without the use of a high temperature target. Such films can be subsequently patterned by dry reactive ion etching. This however sets constraints on possible materials combinations in hybrid structures \cite{Angers08} and is incompatible with connecting niobium to fragile material such as graphene, carbon nanotubes or thin sheets of topological insulators. Sputtering through a mask (resist or mechanical) is also possible \cite{Pallecchi08,Frielinghaus10, Rickhaus12, Komatsu12} but, due to the lack of directionality of the deposition, leads to poorly defined edges. This can cause a gradual loss of the superconducting properties over a broad transition region.

\begin{figure}[t]
\begin{center}
\includegraphics[width=0.7\columnwidth,keepaspectratio]{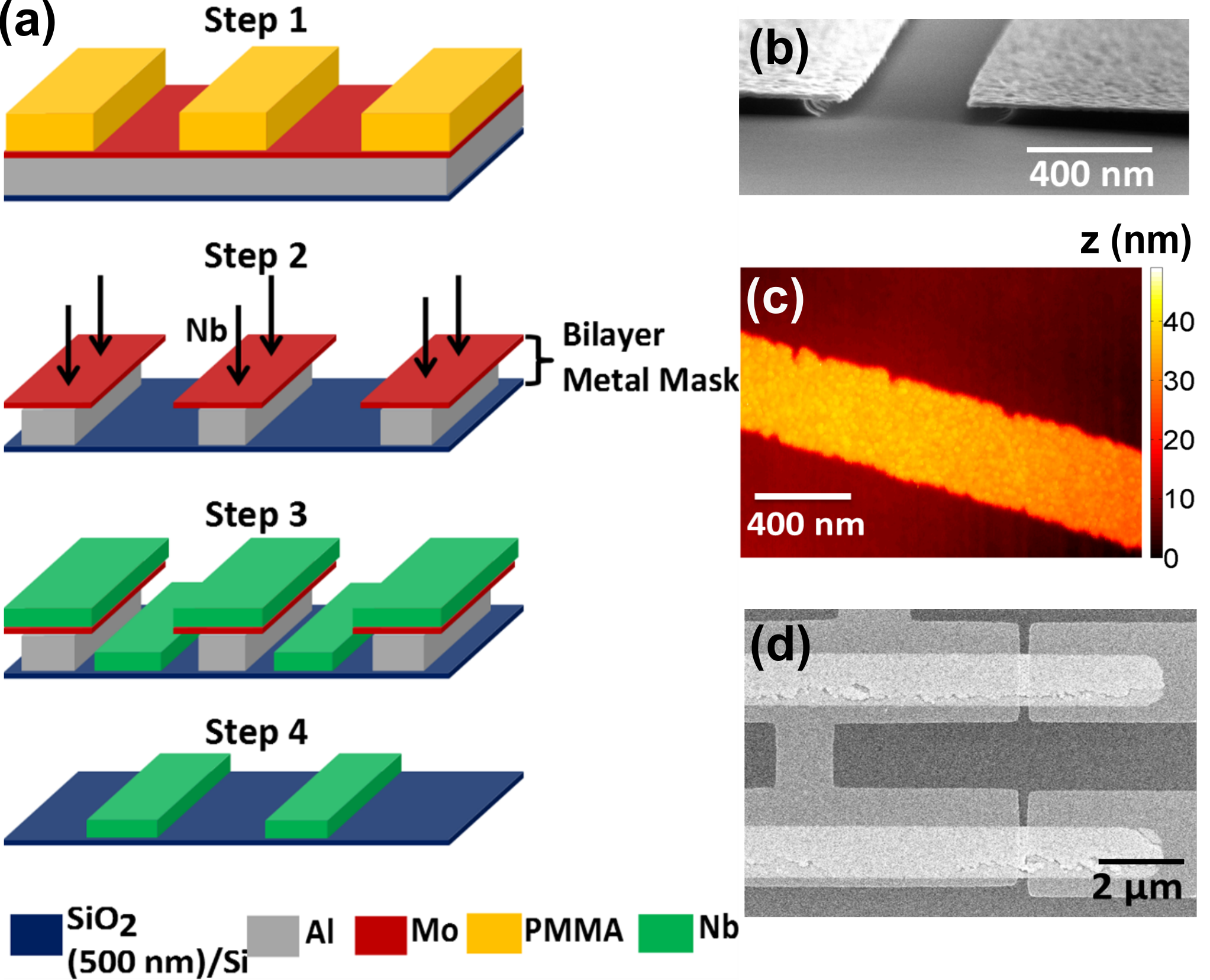}
%\vspace{-6cm}
\caption{(a) Process for a Nb device fabrication. From the top: starting with an Al/Mo (200 nm/40 nm) multilayer, a thin layer of photoresist PMMA (350 nm) is spin coated and patterned using e-beam lithography. In step 2, the Mo layer is patterned using reactive ion etching. This is followed by the wet etching of Al with a basic solution (see text), leaving the Mo top layer free-standing over about 200 nm from the openings. The Nb layer is then evaporated in step 3. In the last step, the remaining Al is dissolved, leaving the sole Nb devices on the substrate. (b) Scanning electron micrograph of the Al/Mo bilayer mask for the wires prior to Nb evaporation. (c) Atomic force micrograph of a small portion of a 400 nm wide Nb line. (d) Scanning electron micrograph of a niobium-gold-niobium proximity SQUID (device C) with 210 nm spacing between the Nb contacts.}
\label{charac}
\end{center}
\end{figure}

Suspended masks made of Si, Ge or SiN, supported by thermostable resists like poly-ethersulfones have further been developed \cite{Dubos00, Hoss02}. Such processes provided Nb sub-micron structures with a critical temperature up to 8 K, including hybrid Josephson junctions with a high Nb-Cu interface transparency \cite{Dubos01}. The use of these particular resists is nevertheless cumbersome; for instance, the ambient hygrometry has to be controlled during resist spinning. Non-organic evaporation stencil masks based on a suspended bilayer of Si$_3$N$_4$ and SiO$_2$ have been developed in the past \cite{Hoss99}; lift-off and integration of arbitrary non-superconductors (graphene etc.) are yet again impossible.
%But the fact that the  Si$_3$N$_4$ layer cannot be lifted off later limits the applications of this process especially to make devices where a local study of the interfaces with scanning tunneling microscopy (STM) would be desirable. Also, contacting arbitrary materials like graphene or carbon natotubes is not possible with this technique. In that case, the final lift-off can be difficult to achieve.
Finally, the possibility of using metallic bilayer masks has been demonstrated in the past \cite{Howard79}. This method was in particular used to pattern large niobium devices with dimensions in the upper 100 $\mu$m range \cite{deWaal81}.

We report here a new and simple method for lithographic patterning of extremely narrow and well-defined Nb nanostructures with a high critical temperature. Our approach relies on replacing the traditional organic resist mask by a fully metallic one. We take advantage of the high wet etch rates of aluminum, with respect to most other metals, by high pH chemicals used as organic resist developers. We demonstrate the fabrication and good operation of SQUIDs based on hybrid Josephson junctions.

%\bigskip
\begin{figure}[t]
\begin{center}
\includegraphics[width=0.55\columnwidth,keepaspectratio]{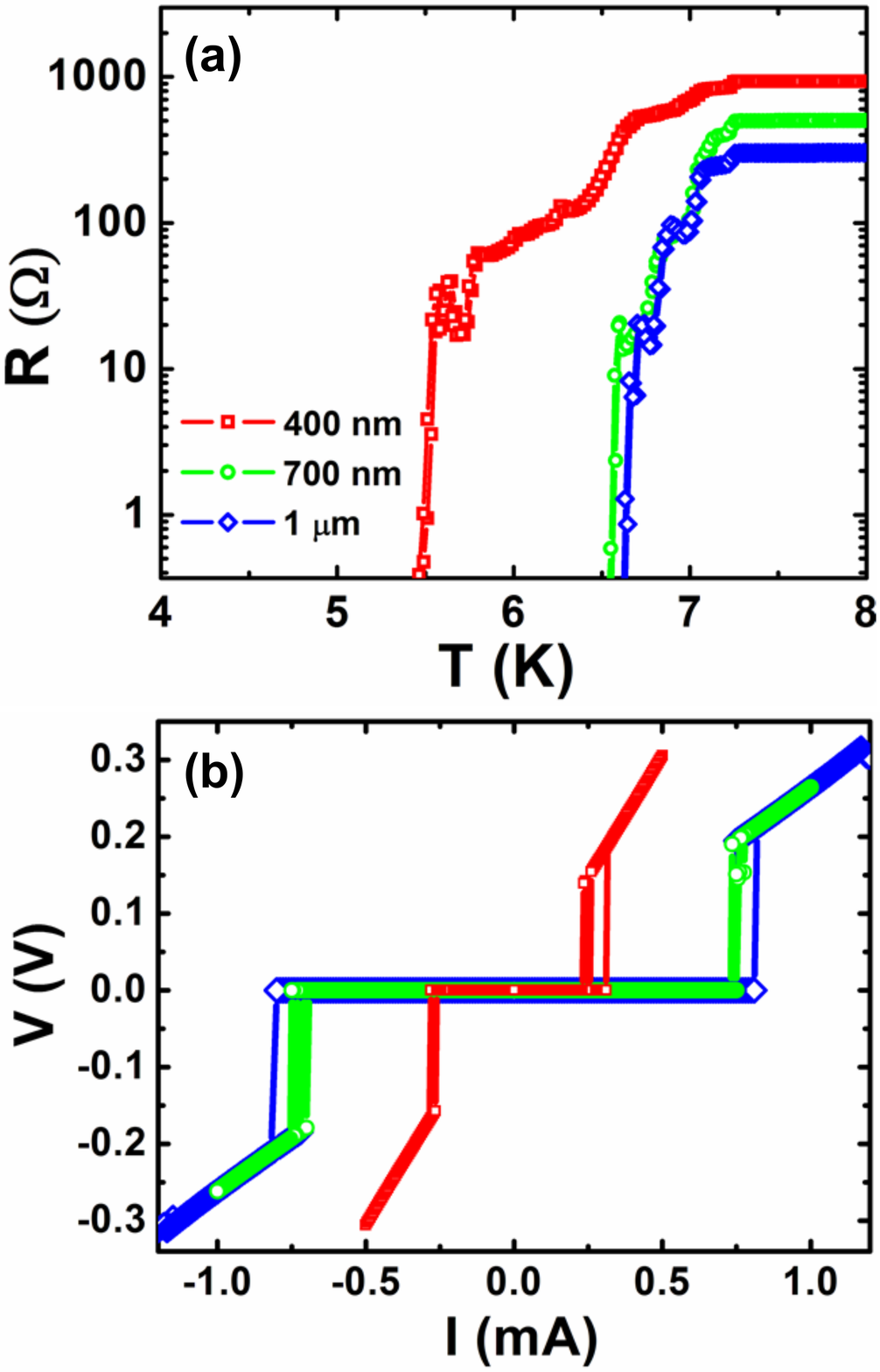}
%\vspace{-0.5cm}
\caption{(a) Resistance versus temperature of three niobium lines of different widths ($I_{bias}=5\, \mu$A). (b) Voltage versus current characteristics of the same samples measured at 4.2 K. These display hysteresis (as high as 32\% for the 400 nm line) as well as random fluctuations of the measured resistance at $I\approx I_c$.}
%(c) Resistance versus magnetic field of the two thinner wires measured at 4.2 K ($I_{bias} = 5\, \mu$A).
\label{wires}
\end{center}
\end{figure}

\section{Fabrication process and submicron lines characterization}

We start by depositing an Al/Mo (200 nm/40 nm) bilayer on a Si/SiO$_2$ substrate. Electron-beam lithography is performed on an organic resist (PMMA) spun on top of the metal bilayer, see Fig. \ref{charac}a. In the developed regions, the Mo layer is removed by a reactive ion etch with a 20 W SF$_6$ plasma for 2 minutes. The resist is afterwards removed by 10 minutes of 50 W O$_2$ plasma etch. In the next step, a wet etch in a basic solution (pH $\approx$ 13) of MF-26A \cite{MF26} etches Al isotropically through the openings in the Mo layer, creating undercuts of about 200 nm after a 90 s time. This leaves a locally suspended Mo mask on top of the over-etched aluminum supporting layer, see Fig. \ref{charac}b,c. This sequence of steps ensures vanishing organic contamination of deposited materials.

A niobium thin film is then e-gun evaporated in an ultra-high vacuum chamber. The metallic Al/Mo mask is finally removed by a wet etch in MF-319 \cite{MF319} for about 45 minutes. Including regularly spaced extra holes in the initial lithographic pattern allows the etchant to start lifting the metal mask from a larger number of starting points. The related minimization of this wet-etch time is important because beyond 2 hours we observe MF-319 starting to etch the Si/SiO$_2$ substrate as well as the metallic layers (Nb, Au,...) present in the device.

As a first test, we have prepared Nb narrow lines using the technique described above. In these samples, a 10 nm Ti layer was deposited in situ prior to Nb, as to improve adhesion and mimic the frequently used approach to contact a superconductor to novel low-dimensional materials \cite{Jarillo06, Heersche07, Sacepe2011}. The presence of Ti however contributes to reducing the $T_c$ of the wires. Atomic force microscopy (AFM) inspection of the Nb-wires obtained (Fig. \ref{charac}c) shows that the edges are extremely sharp and well-defined. The 3$\sigma$ (standard deviations) line edge roughness of the wires is 10 nm, on the order of the grain size of in both the Nb film and the Mo mask. Line edge smearing is inferior to the lateral resolution of a standard AFM tip. The  roughness of the Nb surface is 1.1 nm.

We have performed transport measurements on these samples in a variable temperature (2-300 K) cryostat, using a \mbox{d.c.} 4-probe configuration. Figure \ref{wires}a shows the temperature dependence of the low temperature resistance R of three devices, based on lines of length 30 $\mu$m, thickness 30 nm and widths 400, 700 and 1000 nm respectively. The devices display stepwise decrease of resistance below T= 7.3 K. Since the entire on-chip structure is made of niobium, we attribute these steps to transitions to the superconducting state of wire elements of decreasing width in series. The small resistance steps in the low resistance region can also be related to pinning and depinning of residual vortices. The rapid drop to $R < 1 \Omega$ corresponds to the critical temperature $T_c$ of the narrowest part of each circuit. While the critical temperature $T_c$ of the 400 nm wide line is considerably reduced, it is hardly affected for widths above 700 nm. The $I(V)$ characteristics at 4.2 K show a marked resistive transition and some hysteresis. The latter is due to Joule heating in the Nb wire once in the normal state, which elevates the local electronic temperature compared to the bath temperature. As a result the critical current is reduced, which creates an hysteresis in the current-voltage characteristics. Thermal hysteresis is enhanced in narrower structures.

\section{Fabrication and study of SQUIDs}

We have further fabricated Nb-Au-Nb proximity SQUIDs (Figure \ref{charac}d) using a two-step process. We first patterned two long parallel lines of gold, 1 $\mu$m wide and 30 nm thick, by conventional e-beam lithographic techniques. The metallic mask method described above is then used to pattern Nb proximity junctions on top of these. We deposited a 50 nm thick Nb layer without sticking layer. Over the proximity junction, the metal mask is free-hanging, as to form junctions as short as 200 nm. The device parameters of the three tested devices, labelled A, B and C, are summarized in Table \ref{table}.

In a hybrid Josephson junction made of a normal metal bridging two superconducting electrodes, as a Au wire between two Nb electrodes here, the length scale for inducing superconductivity in the normal metal is set by the normal metal thermal length $L_T=\sqrt{\hbar D/ 2 \pi k_B T}$. Here, $D$ is the electronic diffusion coefficient in the normal metal. $L_T$ is estimated to be about 180 nm at 135 mK while the junction lengths $L$ are in the range 200 -- 400 nm.

\begin{figure}[t]
\begin{center}
\includegraphics[width=0.7\columnwidth,keepaspectratio]{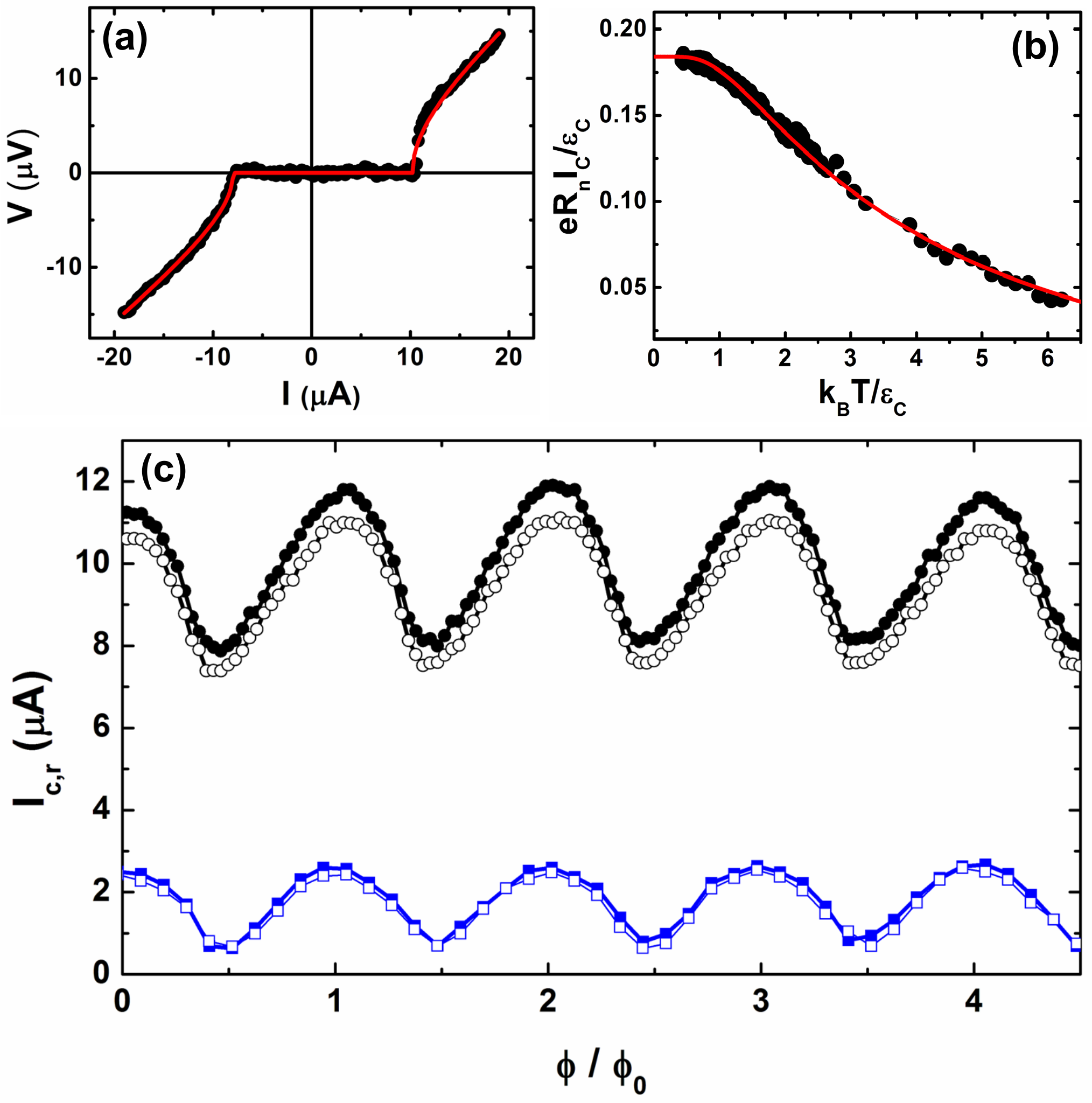}
%\vspace{-5cm}
\caption{(a) $I(V)$ characteristic of device C (circles) at T = 135 mK and B = 0 T. The line is a fit using the over-damped junction theoretical I-V relation, see text. (b) Same device critical current versus temperature. Quantities are normalized to the adjusted Thouless energy $\epsilon_c$ = 36.7 $\mu$eV. The line is a fit by Eq. (\ref{2}). (c) Critical current (full symbols) and retrapping current (hollow symbols) oscillations versus applied magnetic field flux in device B (squares, T = 125 mK) and device C (bullets, T = 135 mK).}
\label{squids}
\end{center}
\end{figure}

Transport properties of the Nb-Au-Nb proximity SQUIDS were measured using a four-probe \mbox{d.c.} current bias scheme inside a dilution refrigerator. Each electrical lead to the samples was thoroughly filtered by individual 2 m long lossy coaxial lines thermalized at the cryostat base temperature. Fig. \ref{squids}a shows the $V(I)$ characteristics of SQUID C at zero magnetic field and 135 mK. A critical current $I_c$ is observed in the two shorter Nb-Au-Nb SQUIDs B and C, while none is observed in A. 

\begin{table*}[ht]
\centering
\setlength{\tabcolsep}{1pt}
\begin{tabular}{c c c c c c c c c}
\hline\hline
  SQUID & $L$(nm) & W($\mu$m) & $R_N$($\Omega$) & $\rho_N$($\mu\Omega$.cm) & $D$(cm$^2/s$) & $L_{T}$(nm) & $I_c$($\mu$A) & $\epsilon_c$($\mu$eV) \\ [0.5ex]
  \hline
  %A & 640 & 0.98 & 2.94 & 13.5 & 29 & 160 &  -  & 4.63 \\
  A & 430 & 0.98 & 1.99 & 13.6 & 29 & 160 &  -  & 10.2 \\
  B & 340 & 1.00 & 1.04 & 9.20 & 42 & 190 & 2.7 & 24.1 \\
    C & 210 & 0.98 & 0.89 & 12.4 & 31 & 170 & 11.8 & 46.7 \\
  \hline\hline
  \end{tabular}
  \caption{Device parameters of the Nb-Au-Nb SQUIDs. $L$ is the geometrical length of the normal weak link (uncovered gold line). $R_N$ is the normal state resistance of the weak link only, measured at 4.2 K, $\rho_N$ the corresponding resistivity and $D$ the diffusion coefficient. $L_{T}$ is the calculated thermal length at T = 135 mK. $I_c$ is the maximum critical current measured at 135 mK. $\epsilon_c = \hbar D/L^2$ is the Thouless energy assuming the weak link length is $L$.}
\label{table}
\end{table*}

As proximity Josephson junctions have a vanishing capacitance, they are over-damped. The $V(I)$ characteristic is thus expected to follow $V=R_n\sqrt{I^2-I_c^2}$ for $|I|> I_c$ \cite{Tinkham}. The characteristics fit rather well this prediction, see Fig. \ref{squids}a. A finite hysteresis is nevertheless observed for the highest $I_c$ SQUID (device C) at the lowest temperatures. While sweeping the current down from large values $|I|> I_c$, the SQUID C turns non-resistive at a retrapping current $I_r \leq I_c$. This hysteresis is known to be of thermal origin \cite{Courtois08}: the Joule heat dissipated in the normal metal elevates the electronic temperature with respect to the phonon temperature as the electron-phonon coupling is the bottleneck for the electron thermalization to the bath. Retrapping happens when the bias current becomes of the order of the critical current at the current electronic temperature.

Our samples are in the long junction limit defined as a normal metal length larger than the superconductor coherence length: $L \gg \xi_s\approx 30$ nm. In this case, the relevant energy scale for the superconducting proximity effect is the Thouless energy $\epsilon_c=\hbar D/L^2$, which is much smaller than the energy gap $\Delta$. The evolution of critical current with temperature in a normal metal weak link at arbitrary temperatures can be understood by solving the Usadel equations. 
%In the high temperature limit, i.e. $k_BT>5\epsilon_C$, this can be simplified to give the following solution for $I_C$ at temperature T  \cite{Dubos01}:
%\begin{equation}
%\frac{eR_nI_c}{\epsilon_c }=\frac{32}{3+2\sqrt{2}}\, \left(\frac{2\pi k_BT}{\epsilon_c} \right)^{3/2}\exp{\left(-\sqrt{\frac{2\pi k_BT}{\epsilon_c}}\right)}.
%\label{1l}
%\end{equation}
%whereas 
At low temperatures and  in the long junction limit ($\Delta/\epsilon_C\rightarrow\infty$), the numerical solution to the Usadel equation can be approximated by \cite{Dubos01}
\begin{equation}
\frac{eR_nI_c}{\epsilon_c }=\eta \, a \left[1-b\, \exp{\left(-\frac{a}{3.2}\frac{\epsilon_c}{k_BT}\right)}\right].
\label{2}
\end{equation}
where a = 10.82,  b = 1.30 and $\eta$ = 1 for perfectly transparent S-N interfaces. 
%In fact, in the range $3 < k_BT/\epsilon_C < 8$ the two above expressions  Eq. \ref{1l} and  Eq. \ref{2} describing the critical current behavior with temperature are found to be numerically equivalent. Hence Eq.\ref{2} has been used to fit the data of $I_C$ vs T for Squid C over the temperature range 182 mK to 2.7 K with $\epsilon_C$ and $\eta$ as the fitting parameters. 
From the fit to the experimental data of SQUID C, shown in Figure \ref{squids}b, values of $\epsilon_c$ = 36.7 $\mu$eV 
%implies that the above temperature range corresponds to the range $0.4 < k_BT/\epsilon_C < 6.2$ where Eq. \ref{2} is perfectly valid. 
and $\eta \approx 0.017$ are found. The estimate of $\epsilon_c $ agrees well with the value found from geometrical arguments, see Table \ref{table}. The minor discrepancy can be understood as an effective junction length $\tilde L=\sqrt{\hbar D/\epsilon_c}$ = 236 nm slightly longer than the geometrical value \cite{Dubos01}. The reduced value of $\eta$ reflects a lower than ideal $I_c$, which is attributed to an imperfect transmission at the Nb-Au interface. Argon plasma of the Au structures prior to Nb deposition should lead to improved contact transparencies. 

Finally, we have measured SQUID B and C critical current variation with the magnetic field applied perpendicular to the loop. Figure \ref{squids}c shows data as a function of the magnetic flux $\phi = B.S$, where $S$ = 37.6 $\mu m^2$ for SQUID B and 19.2 $\mu m^2$ for SQUID C is the geometric loop area. The critical current follows the expected $\phi_0$ periodicity, where $\phi_0$ is the magnetic flux quantum. At a bath temperature T = 135 mK, the modulation depth of $I_c$ reaches 80 \% in sample B. The retrapping current follows the same behavior. The extremal values of  $I_c$ upon flux modulation reflect the sum and the difference respectively of the critical currents of each of the two Josephson junctions in parallel. The obtained flux sensitivity is about 2 $\mu V/\phi_0$. This performance is within the expected range for hybrid SQUIDs that are characterized by a rather low normal state impedance.

\section{Conclusion}

We have devised a new method for the nano-fabrication of superconducting Nb structures using an all-metal mask. The method is easy to implement and uses only standard clean room chemicals and methods. It allows producing superconducting nano-devices with a high critical temperature and dimensions down to the 100 nm-scale. Its application to the realization of sensitive SQUIDS with little hysteresis is demonstrated. Applications of this technique will allow probing Josephson physics in exciting novel material systems such as single crystal nanowires \cite{Wang2009}, ferromagnetic nanowires \cite{Wang2010}, semiconducting nanowires \cite{Doh2005} and graphene \cite{Jarillo06}. This method can also be used for the fabrication of tunnel junction-based devices like single-electron transistors.

\section*{Acknowledgments}
This work is funded by the EU Marie Curie ITN Q-NET. A. F. acknowledges funding from Grenoble INP BQR. Samples were fabricated at Nanofab platform at CNRS Grenoble. The authors want to thank T. Fournier, T. Crozes, S. Martin, L. Pascal, J. Marcus and F. Gay for help and discussions about fabrication and measurement issues.


\section*{References}

\begin{thebibliography}{30}

\bibitem{Vion2002} Vion D, Aassime A, Cottet A, Joyez P, Pothier H, Urbina C, Esteve D and Devoret M H 2002 Manipulating the quantum state of an electrical circuit \emph{Science} \textbf{296}, 886
\bibitem{Wallraff04} Wallraff A, Schuster D I, Blais A, Frunzio L, Huang R-S, Majer J, Kumar S, Girvin S M, Schoelkopf R J 2004 Strong coupling of a single photon to a superconducting qubit using circuit quantum electrodynamics  \emph{Nature} \textbf{431} 162
\bibitem{Tinkham} Tinkham M, Introduction to Superconductivity, second edition, Dover Publications 2004
\bibitem{Spathis11} Spathis P, Biswas S, Roddaro S, Sorba L, Giazotto F and Beltram F 2011 Carbon nanotube josephson junctions with Nb contacts \NT \textbf{22}, 105201
\bibitem{Wei2008} Wei J, Olaya D, Karasik B S, Pereverzev S V,  Sergeev A V and Gershenson M E 2008 Ultrasensitive hot-electron nanobolometers for terahertz astrophysics \emph{Nature Nanotech.} \textbf{3}, 496
\bibitem{Ohnishi08} {Ohnishi} K, {Kimura} T, and {Otani} Y 2008 Improvement of Superconductive Properties of Mesoscopic Nb Wires by Ti Passivation Layers \emph{Appl. Phys. Express} \textbf{1}, 021701 
\bibitem{Kim01} Kim N,  Hansen K, Toppari J, Suppula T and Pekola J 2002 Fabrication of mesoscopic superconducting Nb wires using conventional
  electron-beam lithographic techniques \emph{J. Vac. Sci. Technol. B} \textbf{20}, 386
\bibitem{Angers08} Angers  L, Chiodi F,  Montambaux G,  Ferrier M, Gu\'eron S, Bouchiat H and Cuevas J C 2008 Proximity dc squids in the long junction limit \emph{Phys. Rev. B} \textbf{77}, 165408 
\bibitem{Pallecchi08} Pallecchi E, Gaass M, Ryndyk D A and  Strunk Ch 2008 Carbon nanotube Josephson junctions with Nb contacts \emph{Appl. Phys. Lett.} \textbf{93}, 072501
\bibitem{Frielinghaus10} Frielinghaus R, Batov I E, Weides M, Kohlstedt H, Calarco R and Sch\"{a}pers Th 2010 Josephson supercurrent in Nb/InN-nanowire/Nb junctions \emph{Appl. Phys. Lett.} \textbf{96}, 132504 
\bibitem{Rickhaus12} Rickhaus P, Weiss M,  Marot L and Sch\"{o}nenberger C 2012 Quantum hall effect in graphene with superconducting electrodes \emph{ Nano Lett.} \textbf{12}, 1942 
\bibitem{Komatsu12} Komatsu K, Li C, Autier-Laurent S, Bouchiat H and Gu\'eron S 2012 Superconducting proximity effect in long superconductor/graphene/superconductor junctions: From specular andreev  reflection at zero field to the quantum hall regime \emph{Phys. Rev. B} \textbf{86}, 115412 
\bibitem{Dubos00} Dubos P, Charlat P, Crozes Th, Paniez P and Pannetier B 2000 Thermostable trilayer resist for niobium lift-off \emph{J. Vac. Sci. Technol. B} {\bf 18}, 122 
\bibitem{Hoss02} Hoss T, Strunk C, S{$ \rm \ddot{u}$}rgers C, and Sch{$ \rm \ddot{o}$}nenberger C 2002 UHV compatible nanostructuring technique for mesoscopic hybrid devices: application to superconductor/ferromagnet josephson contacts \emph{Physica E} \textbf{14}, 341 
\bibitem{Dubos01} Dubos P, Courtois H, Pannetier B, Wilhelm F K, Zaikin A D and Sch{$ \rm \ddot{o}$}n G 2004 Josephson critical current in a long mesoscopic s-n-s junction \emph{Phys. Rev. B} \textbf{63}, 064502 
\bibitem{Hoss99} Hoss T, Strunk C, and Sch{$ \rm \ddot{o}$}nenberger C 1999 Nonorganic evaporation mask for superconducting devices \emph{Microelec. Engineering} \textbf{46}, 149
\bibitem{Howard79} Howard R E 1978 A refractory lift-off process with applications to high-T$_c$ superconducting circuits \emph{Appl. Phys. Lett.} \textbf{33}, 1034
\bibitem{deWaal81} de Waal V J, van den Hamer P, Mooij J E, Klapwijk T M 1981 Very low noise all-niobium SQUIDS \emph{IEEE Transactions on Magnetics} \textbf{17}, 858 
\bibitem{MF26} Microposit MF-26A Developer, Shipley Europe Ltd.
\bibitem{MF319} Microposit MF-319 Developer, Shipley Europe Ltd.
\bibitem{Jarillo06} Jarillo-Herrero P, van Dam J A and Kouwenhoven L P 2006 Quantum supercurrent transistors in carbon nanotubes \emph{Nature} \textbf{439}, 953
\bibitem{Heersche07} Heersche H B, Jarillo-Herrero P, Oostinga J B, Vandersypen L M K and Morpurgo A F 2007 Bipolar supercurrent in graphene \emph{Nature} \textbf{446}, 56
\bibitem{Sacepe2011} Sac\'ep\'e B, Oostinga J B, Li J, Ubaldini A, Couto N J G, Giannini E and Morpurgo A F 2011 Gate-tuned normal and superconducting transport at the surface of a topological insulator \emph{Nature Comm.} \textbf{2}, 575
\bibitem{Courtois08} Courtois H, Meschke M, Peltonen J T and Pekola J P 2008 Origin of hysteresis in a proximity josephson junction \emph{Phys. Rev. Lett.} \textbf{101}, 067002 
\bibitem{Wang2009} Wang J, Shi C, Tian M, Zhang Q, Kumar N, Jain J K, Mallouk T E, Chan M H W 2009 Proximity-induced superconductivity in nanowires: Minigap state and differential magnetoresistance oscillations \emph{Phys. Rev. Lett.} \textbf{102}, 247003
\bibitem{Wang2010} Wang J, Singh M, Tian M, Kumar N, Liu B, Shi C, Jain J K, Samarth N, Mallouk T E, Chan M H W  2010 Interplay between superconductivity and ferromagnetism in crystalline nanowires \emph{Nature Physics} \textbf{439}, 953
\bibitem{Doh2005} Doh Y J, van Dam J A, Roest A L, Bakkers E P A M, Kouwenhoven L P, de Franceschi S 2005 Tunable Supercurrent Through Semiconductor Nanowires \emph{Science} \textbf{309}, 272

\end{thebibliography}
\end{document}